# On the Landauer formula in interfacial thermal transport

Jinghang Dai and Zhiting Tian[*]

*Sibley School of Mechanical and Aerospace Engineering, Cornell University, NY 14853, USA*

**Abstract**

In this commentary, we clarify that the Landauer formula is not limited to the phonon gas model. It is fundamentally more general and applies to both particle- and wave-based descriptions of phonons, provided the transmission function is well defined. In the harmonic regime, the phonon transmission function and the resulting Landauer expression for heat current are exact. They can be rigorously derived using the atomistic Green's function method, which treats phonons as waves and does not require phonon dispersion in the interface region. In short, the Landauer framework remains valid for ideal, disordered, and defective interfaces, as long as an appropriate transmission function is used.

*Address all correspondence to Zhiting Tian's E-mail: zhiting@cornell.edu

**Introduction**

The Landauer formalism treats transport as a transmission process [1]. Although widely used in electronic transport [2], its validity and applicability in interfacial thermal transport have been less clear. A common misconception is that the Landauer formula relies on the phonon gas model [3,4], which would render it inapplicable when phonon dispersion and group velocities are ill-defined. In reality, the Landauer approach is far more general and remains valid even when phonons are treated as waves. For example, the Landauer formula can be rigorously derived using atomistic Green's function (AGF) [5] or quantum Langevin equations [6], neither of which requires phonon dispersion in the interface region.

In this commentary, we use AGF to demonstrate that the Landauer formula is not rooted in the phonon gas model. The AGF approach has been used to study all types of interfacial heat conduction [7–10]. A typical model used for interfacial thermal transport is depicted in Fig. 1. The system is divided into three parts: two semi-infinite leads (left and right) and the central region, which includes the physical interface. The leads are set to be crystalline structures and are in thermal equilibrium with their respective temperature baths. One can define a supercell for an amorphous structure in the leads to include the disorder to a certain extent, but the contribution of some long-wavelength modes would be neglected. We thus limit our discussions to crystalline leads. Regardless, we can explore a wide variety of interfaces because the central region does not need to exhibit well-defined phonon dispersion. The transmission function is valid as long as one can obtain the force constants of the system.

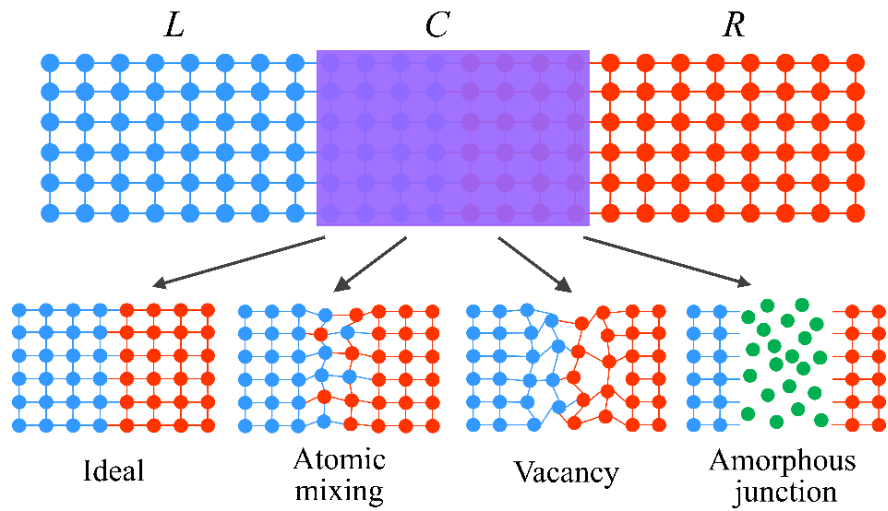

*Fig. 1 A general model of interfacial thermal transport; L stands for the left lead, C stands for the central region, and R stands for the right lead. The central region can be various structures such as ideal interfaces, defective interfaces, and amorphous junctions.*

**Discussion**

To be complete, we first show the derivation of the Landauer formula under the phonon gas model. As mentioned in some previous work [11], the heat current inflow across a 3D structure from either side is:

$$j_{L(R)} = A_s \times \int_0^\infty d\omega \frac{1}{2}\frac{v(\omega)}{2} DOS_{L(R)}(\omega)\tau_{L(R)}(\omega)\hbar\omega f_{B.E.}(\omega, T_{L(R)}) \tag{1}$$

where $A_s$ is the cross-sectional area. Based on the phonon gas model, phonons are treated as quasi-particles, namely gas particles that carry energy $\hbar\omega$ and the group velocity as $v(\omega)$, where $\omega$ is the phonon frequency. $\tau_{L(R)}$ and $DOS_{L(R)}$ are the transmittance (transmission probability of a given mode) and density of states of left (or right) lead, while $f_{B.E.}(\omega, T_{L(R)}) = 1/(e^{\hbar\omega/k_B T_{L(R)}} - 1)$ is the Bose-Einstein distribution. Suppose we introduce the number of phonon modes $M(\omega)$, which is given as $M(\omega) = \pi A_s \frac{v(\omega)}{2} DOS_{L(R)}(\omega)$ for 3D systems [11], and define the phonon transmission function as $\Xi(\omega) = M(\omega)\tau_{L(R)}(\omega)$, which describes how many modes are transmitted and is the same for both leads guaranteed by detailed balance, then the heat current from left (or right) lead to right (or left) lead will be:

$$j_{L(R)} = \frac{1}{2\pi}\int_0^\infty \hbar\omega f_{B.E.}(\omega, T_{L(R)})\Xi(\omega)\, d\omega \tag{2}$$

If the temperature of the left lead is higher than the right lead, the net heat current across the interface from the Landauer formula for phonon transport becomes (from high temperature to low temperature):

$$J = j_L - j_R = \frac{1}{2\pi}\int_0^\infty \hbar\omega [f_{B.E.}(\omega, T_L) - f_{B.E.}(\omega, T_R)]\Xi(\omega)d\omega \tag{3}$$

We then derive the Landauer formula using the wave nature of phonons to show that the phonon gas model is by no means the prerequisite for the Landauer formula to be valid. The following derivation is primarily based on previous papers [12] of AGF from Fisher's group. Instead of focusing on the algebra or its implementation for real interfaces, we interpret each step and its conditions to demonstrate that the phonon gas model is not the basis of the Landauer formula. Here, we conduct all the derivations in the harmonic regime. When the temperature is low enough that the anharmonicity is negligible, such an assumption is appreciated in general. To include anharmonicity, please refer to Mingo's original paper on 1D structures [13] and our recent development on 3D structures [14,15].

Let us consider a vector of atomic displacements $\{u_i\}$ (not necessarily an eigen solution). The total energy of the $i$th degree of freedom will be:

$$E_i = \frac{1}{4}\sum_j (u_i^* k_{ij} u_j + u_j^* k_{ji} u_i) + \frac{M_i}{2}\dot{u}_i^*\dot{u}_i \tag{4}$$

where $k_{ij}$ is the force constant between the $i$th and $j$th degree of freedom. By taking the derivative over time for Eq. (4) and using Newton's law $m_i \ddot{u}_i = -\sum_j k_{ij} u_j$, the heat current between any two degrees of freedom can be represented by:

$$J_{ij} = \frac{1}{4}\left(u_i^* k_{ij} \dot{u}_j + \dot{u}_j^* k_{ji} u_i - u_j^* k_{ji} \dot{u}_i - \dot{u}_i^* k_{ij} u_j\right) \tag{5}$$

By using the complex wave expression $u_i = \phi_i e^{-i\omega t}/\sqrt{M_i}$ (not necessarily a plane wave), we obtain a simplified heat current equation:

$$J_{ij} = \frac{\omega}{2i}\left(\phi_i^* \tilde{k}_{ij} \phi_j - \phi_j^* \tilde{k}_{ji} \phi_i\right) \tag{6}$$

where $\tilde{k}_{ij}$ is the element force constant matrix divided by mass or named as the dynamical matrix, which is defined as $\tilde{k}_{ij} = k_{ij}/\sqrt{M_i M_j}$. Importantly, Eq. (6) is a generally valid equation for all complex waves of the form $u_i = \phi_i e^{-i\omega t}/\sqrt{M_i}$.

In order to compute the energy transport in the system defined as Fig. 1, we construct the dynamical matrices for the central region and leads. First, we write the dynamical equations for the uncoupled leads:

$$\left(\omega^2 I - H_{L(R)}\right)\Phi_{L(R)} = 0 \tag{7}$$

where $H_{L(R)}$ refers to the dynamical matrix of the left (or right) lead, and $\Phi_{L(R)}$ are the vibrational degrees of freedom multiplied by the square root of the atomic mass for the uncoupled leads defined as $\Phi_{L(R),i} = u_{L(R),i} \times \sqrt{M_i}$. Next, once we couple the leads and the central region together, the dynamical equations for the whole system [10] become:

$$\begin{pmatrix} \omega^2 I - H_L & -D_L^\dagger & 0 \\ -D_L & \omega^2 I - H_C & -D_R \\ 0 & -D_R^\dagger & \omega^2 I - H_R \end{pmatrix} \begin{pmatrix} \Phi_L + \Delta_L \\ \Psi \\ \Phi_R + \Delta_R \end{pmatrix} = 0 \tag{8}$$

where $D_{L(R)}$ refers to the dynamical matrix connecting the left (or right) lead and the central region. $\Delta_{L(R)}$ represents changes in the leads after the coupling, and $\Psi$ is the vibrational degrees of freedom multiplied by the square root of the atomic mass for the central region. The notation "$\dagger$" means conjugate transpose. It is worth mentioning that even when the central region is an amorphous material and proper phonon dispersion cannot be defined, we can still calculate the dynamical matrix $H_C$ since we form the dynamical equations in real space (i.e., at the Gamma point only). There is no requirement for translational symmetry along the heat conduction direction in the central region because well-defined wave vectors are not necessary for eigen solutions. What we only need to do here is to solve for the eigenvalues and eigenvectors for a given matrix. The eigen solutions do not even need to be plane waves.

Finally, we can solve for the eigen solutions $\Delta_L$, $\Delta_R$, and $\Psi$ as long as all the dynamical matrices can be defined. As a special attribute of Green's functions, we can represent these eigen solutions as:

$$\Delta_L = g_L D_L^\dagger \Psi \tag{9}$$

$$\Delta_R = g_R D_R^\dagger \Psi \tag{10}$$

$$\Psi = G_C(D_L\Phi_L + D_R\Phi_R) \tag{11}$$

where the surface Green's function for the leads is defined as $g_{L(R)} = \left((\omega^2 + i\eta)I - H_{L(R)}\right)^{-1}$ ($i$ is the imaginary unit, and $\eta$ is a positive infinitesimal number) and the central Green's function is defined as $G_C = \left(\omega^2 I - H_C - D_L g_L D_L^\dagger - D_R g_R D_R^\dagger\right)^{-1}$.

With the help of the Green's function, one can compute the heat current density for the interface without approximation. To begin, we calculate the heat current between the left lead and the central region. By inserting the eigen solutions from Eq. (8) into Eq. (6), we have:

$$J_{L\to C} = \frac{\omega}{2i}[\Psi^\dagger D_L(\Phi_L + \Delta_L) - (\Phi_L + \Delta_L)^\dagger D_L^\dagger \Psi] \tag{12}$$

It is worth restating for Eq. (12) instead of considering phonon particles to carry energy, we can represent the heat current with the solutions of the eigenproblem.

Using Eq. (12), the algebra is straightforward with Eqs. (9), (10), and (11), to obtain the net heat current between the left lead and the central region. Note that Eq. (12) is only for the contribution from one mode of a given eigenstate from the left lead. To calculate the total heat current, we also need to sum over all eigenstates and account for the phonon population in each. For further details, please refer to the Appendix. The final expression is:

$$J_{L\to C-total} = \int_0^\infty \frac{\hbar\omega}{2\pi} \text{Tr}[\Gamma_L G_C \Gamma_R G_C^\dagger](f_{B.E.}(\omega, T_L) - f_{B.E.}(\omega, T_R))d\omega \tag{13}$$

where $f_{B.E.}(\omega, T_{L(R)})$ is the phonon occupation corresponding to a given temperature of the left (or right) lead at eigenstate $\omega^2$, or the Bose-Einstein distribution we mentioned before. $\Gamma_{L(R)}$ is the "escape rate" for the connection between the left (or right) lead and the central region, and the details can be found in Appendix. All the derivations from Eq. (12) to Eq. (13) are simple algebra and do not invoke the phonon gas assumption in the central region. In like manner, we can obtain the net heat current between the right lead and the central region:

$$J_{R\to C-total} = \int_0^\infty \frac{\hbar\omega}{2\pi} \text{Tr}[\Gamma_R G_C \Gamma_L G_C^\dagger](f_{B.E.}(\omega, T_R) - f_{B.E.}(\omega, T_L))d\omega \tag{14}$$

It is easy to verify that $J_{L\to C-total} = -J_{R\to C-total}$. If we define the transmission function as $\Xi(\omega) = \text{Tr}[\Gamma_R G_C \Gamma_L G_C^\dagger]$, we can rewrite net heat current from the left lead to the right lead in the same format as the Landauer formula in Eq. (3):

$$J = \frac{1}{2\pi}\int_0^\infty \hbar\omega\,[f_{B.E.}(\omega, T_L) - f_{B.E.}(\omega, T_R)]\,\Xi(\omega)d\omega \tag{15}$$

For the above derivations, no assumption of the phonon gas model in the central region is required. Within the AGF framework, the spectral transmission function $\Xi(\omega)$ can be computed directly. Since all other quantities are known, the total heat flow $J$ across the interface can be evaluated, and the interfacial thermal conductance is subsequently obtained by normalizing $J$ by the interface cross-sectional area. In fact, under the wave picture, no matter how scattering centers are distributed in the central region, one can always calculate a transmission function which describes the propagation of waves between the two leads [1]. Even with disordered central regions where the phonon dispersion cannot be defined, one can always compute the transmission function, and the total heat current can thus be obtained [9].

**Conclusion**

In summary, the phonon gas model is not the foundation of the Landauer formula; it is simply one special case that satisfies it. Using AGF, we clarified the derivation of the Landauer expression for thermal transport by explicitly accounting for the wave nature of phonons. We aim to emphasize that the Landauer framework is broadly applicable to all types of interfaces, provided the transmission function is well defined. By resolving this misconception, we hope to encourage wider and more confident use of the Landauer approach wherever it is applicable.

**Competing Interest**

On behalf of all authors, the corresponding author states that there is no competing interest.

**Authors' Contribution**

J.D. carried out the derivation and wrote the manuscript draft, Z.T. conceptualized the idea, secured funding, and revised the manuscript.

**Data availability**

The data that support the findings of this study are available within the article and its appendix.

**Funding**

This work was sponsored by the Department of the Navy, Office of Naval Research under ONR Awards No. N00014-18-1-2724 and N00014-22-1-2357.

**Acknowledgment**

We thank Prof. Timothy S. Fisher and Dr. Joseph L. Feldman for their valuable feedback. We thank Prof. Keivan Esfarjani for useful discussions.

**Appendix**

In the Appendix, we will present the mathematical derivations from Eq. (12) to Eq. (13). Here are some useful matrices we will use in the following derivations:

$$\boldsymbol{A} = i(\boldsymbol{G}_C - \boldsymbol{G}_C^\dagger) = \boldsymbol{A}_L + \boldsymbol{A}_R \quad (A1)$$

where $\boldsymbol{A}_{L(R)} = \boldsymbol{G}_C \boldsymbol{\Gamma}_{L(R)} \boldsymbol{G}_C^\dagger$, and $\boldsymbol{\Gamma}_{L(R)} = i\boldsymbol{D}_{L(R)}(\boldsymbol{g}_L - \boldsymbol{g}_L^\dagger)\boldsymbol{D}_{L(R)}^\dagger$. $\boldsymbol{A}_{L(R)}$ is named as "spectral function" for the leads and $\boldsymbol{\Gamma}_{L(R)}$ is the "escape rate" for the connection between the left (or right) lead and the central region.

In general, Eq. (12) can be divided into two terms:

$$J_{L\to C} = J_{L\to C,term1} - J_{L\to C,term2} = \frac{\omega}{2i}[\Psi^\dagger \boldsymbol{D}_L \Phi_L - \Phi_L^\dagger \boldsymbol{D}_L^\dagger \Psi] - \frac{\omega}{2i}[\Psi^\dagger \boldsymbol{D}_L \Delta_L - \Delta_L^\dagger \boldsymbol{D}_L^\dagger \Psi] \quad (A2)$$

We can evaluate the first term of Eq. (A2) using Eqs. (9), (10), (11), (A1):

$$J_{L \to C, term1} = \frac{\omega}{2i}\left[\Psi^\dagger D_L \Phi_L - \Phi_L^\dagger D_L^\dagger \Psi\right]$$

$$= \frac{\omega}{2i}\left[(G_C(D_L\Phi_L + D_R\Phi_R))^\dagger D_L\Phi_L - \Phi_L^\dagger D_L^\dagger G_C(D_L\Phi_L + D_R\Phi_R)\right]$$

$$= \frac{\omega}{2i}\text{Tr}[\Phi_L(\Phi_L^\dagger D_L^\dagger + \Phi_R^\dagger D_R^\dagger)G_C^\dagger D_L - \Phi_L\Phi_L^\dagger D_L^\dagger G_C D_L - \Phi_R\Phi_L^\dagger D_L^\dagger G_C D_R]$$

(A3)

where we use the properties of the trace of the matrix by moving $\Phi_{L(R)}$ to the left side. Additionally, since $\Phi_L$ and $\Phi_R$ are the degrees of freedom vectors of the two uncoupled leads, the inner product will be zero, which can also be proved using Eq. (7). As a result, the first term of Eq. (A2) will be:

$$J_{L \to C, term1} = \frac{\omega}{2i}\text{Tr}[\Phi_L\Phi_L^\dagger D_L^\dagger G_C^\dagger D_L - \Phi_L\Phi_L^\dagger D_L^\dagger G_C D_L]$$

$$= \frac{\omega}{2i}\text{Tr}[D_L\Phi_L\Phi_L^\dagger D_L^\dagger G_C^\dagger - D_L\Phi_L\Phi_L^\dagger D_L^\dagger G_C]$$

$$= \frac{\omega}{2}\text{Tr}[D_L\Phi_L\Phi_L^\dagger D_L^\dagger A] \qquad (A4)$$

where we have used the fact that the trace of any matrix product is independent of the order of multiplication. To acquire the total heat current contributed by the first term, we need to sum Eq. (A4) over the different eigenvectors (denoted by subscript $k$) of the left lead, which will be:

$$J_{L \to C, term1-total} = \frac{1}{2}\sum_k \omega_k f_{B.E.}(\omega_k, T_L)\text{Tr}[D_L\Phi_{L,k}\Phi_{L,k}^\dagger D_L^\dagger A] \qquad (A5)$$

where the Bose-Einstein distribution is the phonon occupation of the state $k$. As can be seen, to formulate the heat current using the Green's function, we need to solve for the normalization condition for phonons at the leads. Based on Eq. (4) and the expression $u_i = \phi_i e^{-i\omega t}/\sqrt{M_i}$, the classical energy for the $i$th degree of freedom with the eigenvalue $\omega_k^2$ (or eigenfrequency $\omega_k$) is:

$$E_{i,k} = \omega_k^2|\phi_{L,i,k}|^2 \qquad (A6)$$

The classical energy of the $i$th degree of freedom over a frequency square range of ($\omega^2$, $\omega^2 + d\omega^2$) is given by:

$$E_i(\omega^2) = \sum_{\omega_k^2 \in (\omega^2, \omega^2+d\omega^2)} \omega_k^2|\phi_{L,i,k}|^2 \qquad (A7)$$

On the other hand, the quantum energy of phonon modes in the left lead over the frequency square range ($\omega^2$, $\omega^2 + d\omega^2$), and associated with the $i$th degree of freedom is given by:

$$E_i(\omega^2) = \hbar\omega DOS_L(\vec{r}_i, \omega^2)d\omega^2 \qquad (A8)$$

where $\hbar\omega$ is the energy for a phonon and the $DOS_L(\vec{r}_i, \omega^2)$ is the local density of states for the $i$th degree of freedom in the left lead. By equating Eq. (A7) and (A8), we can have the following relationship:

$$\sum_{\omega_k^2 \in (\omega^2, \omega^2+d\omega^2)} \omega_k |\phi_{L,i,k}|^2 = \hbar DOS_L(\vec{r}_i, \omega^2)d\omega^2 \tag{A9}$$

Using the well-known fact that the local density of states at the *i*th degree of freedom is given by the *i*th diagonal component of the spectral function of the uncoupled lead [10], we can rewrite Eq. (A9) and extend the respective equations to their corresponding matrix forms:

$$\sum_{\omega_k^2 \in (\omega^2, \omega^2+d\omega^2)} \omega_k \Phi_{L,k}\Phi_{L,k}^\dagger = \hbar \frac{A_L(\omega)}{2\pi} d\omega^2 \tag{A10}$$

In such a small frequency square range, the spectral function $A$ in Eq. (A5) can be considered as a constant matrix. As a result, we can obtain a similar equation to Eq. (A5):

$$\sum_{\omega_k^2 \in (\omega^2, \omega^2+d\omega^2)} \omega_k D_L \Phi_{L,k}\Phi_{L,k}^\dagger D_L^\dagger A_k = \frac{\hbar}{2\pi} D_L A_L(\omega) D_L^\dagger A(\omega) d\omega^2 \tag{A11}$$

Considering the total energy flow by summing over all the possible $k$:

$$\sum_{all\ k} \omega_k D_L \Phi_{L,k}\Phi_{L,k}^\dagger D_L^\dagger A_k = \int_0^\infty \frac{\hbar}{2\pi} D_L A_L(\omega) D_L^\dagger A(\omega) d\omega^2 \tag{A12}$$

By multiplying the factor "$\frac{1}{2} f_{B.E.}(\omega, T_L)$" from Eq. (A5), the trace of the above equation yields:

$$J_{L \to C, term1-total} = \int_0^\infty \frac{\hbar\omega}{2\pi} \text{Tr}[D_L A_L D_L^\dagger A] f_{B.E.}(\omega, T_L) d\omega = \int_0^\infty \frac{\hbar\omega}{2\pi} \text{Tr}[\Gamma_L A] f_{B.E.}(\omega, T_L) d\omega \tag{A13}$$

in which we dropped the frequency notation of each spectral function just for brevity. Similarly, the second term of Eq. (A2) will be:

$$J_{L \to C, term2-total} = \int_0^\infty \frac{\hbar\omega}{2\pi} \big( \text{Tr}[A_L \Gamma_L] f_{B.E.}(\omega, T_L) + \text{Tr}[A_R \Gamma_L] f_{B.E.}(\omega, T_R) \big) d\omega \tag{A14}$$

Therefore, we can express the net heat current from the left lead to the central region as:

$$J_{L \to C-total} = J_{L \to C, term1-total} - J_{L \to C, term2-total}$$

$$= \int_0^\infty \frac{\hbar\omega}{2\pi} \text{Tr}[\Gamma_L G_C \Gamma_R G_C^\dagger] \big( f_{B.E.}(\omega, T_L) - f_{B.E.}(\omega, T_R) \big) d\omega \tag{A15}$$